\documentclass[
10pt,tightenlines, amsmath, amssymb, nofootinbib, prl,twocolumn,
superscriptaddress, showpacs, preprintnumbers, letter]{revtex4}

\usepackage{graphicx}

\newcommand{\beq}{\begin{equation}}
\newcommand{\eeq}{\end{equation}}
\newcommand{\bea}{\begin{eqnarray}}
\newcommand{\eea}{\end{eqnarray}}
\newcommand{\nn}{\nonumber}

\def\pzr{ |p\!\uparrow\rangle }
\def\pzl{ \langle p\!\uparrow\!| }
\def\ppzl{ \langle p'\!\uparrow\!| }
\def\pzn{ \langle p\!\uparrow\!|p\!\uparrow\rangle }

\def\ua{\uparrow}
\def\da{\downarrow}

\newcommand{\Tr}{{\rm Tr}}

\newcommand{\junk}[1]{}

\begin{document}

\title{ Heavy-quark contribution to the proton's magnetic moment}
\author{Xiangdong Ji}
\affiliation{Department of Physics, University of Maryland, College
Park, Maryland 20742, USA} \affiliation{Department of Physics,
Peking University, Beijing, 100871, P. R. China}
\author{D.~Toublan}
\affiliation{Department of Physics, University of Maryland, College
Park, Maryland 20742, USA}
\date{\today}
\begin{abstract}
We study the contribution to the proton's magnetic moment
 from a heavy quark sea in quantum chromodynamics. The heavy quark is
integrated out perturbatively to obtain an effective dimension-6
magnetic moment operator composed of three gluon fields. The leading
contribution to the matrix element in the proton comes from a quadratically
divergent term associated with a light-quark tensor
operator. With an approximate knowledge of the proton's tensor charge,
we conclude that a heavy sea-quark contribution to the proton's magnetic moment is positive in the asymptotic limit. We comment on the
implication of this result for the physical strange quark.
\end{abstract}


\maketitle

{\bf 1.} {\it Introduction.} Naively, the proton is composed of
three valence quarks bound by the strong interaction. However,
quantum chromodynamics (QCD), the theory of strong interactions,
implies a much more complex structure: The proton also contains a
sea of virtual gluons and quark-antiquark pairs.  These virtual
components of the proton have measurable consequences on the proton's
macroscopic properties, such as its mass and spin. In this article,
we study the contribution to the magnetic moment of the proton from
a heavy sea quark with mass $m_Q\gg \Lambda_{\rm QCD}$, the strong
interaction scale.

Our study is motivated by the recent results from a series of
remarkable experiments carried out at MIT-Bates, Jefferson Lab, and
Mainz \cite{Aniol:2005zg}. They all seem to indicate, within
experimental errors, that the strange quark sea makes a positive
contribution to the proton's magnetic moment in the standard
convention that leaves out the quark's electric charge factor.
On the other hand,
most models,
along with some lattice-based calculations, predict a negative
strange quark contribution to the proton's magnetic moment \cite{Beise:2004py}.
Therefore, there seem to be a
contradiction between experimental data and theoretical
results. A resolution of this contradiction might help us to
better understand the influence of the strange quark on the macroscopic properties of the proton.

A direct calculation of the strange quark contribution to the proton's
magnetic moment is a challenging task, mainly because the strange quark is neither
too light nor too heavy compared to
$\Lambda_{\rm QCD}$: Its mass lies just at the border between these two regimes.
This is unfortunate because various successful theoretical methods
have been developed for these two limits only. On the other hand, it might be useful to treat
the strange quark mass as a variable, and to study the
limiting cases where the strange quark is either very light or very heavy.

There is strong evidence that a light sea-quark contributes
negatively to the proton's magnetic moment, although a rigorous
proof does not yet exist. In this article, we focus on the opposite
limit: We consider a heavy strange quark and evaluate its
contribution to the proton's magnetic moment. By integrating out the
heavy quark perturbatively, we obtain an effective magnetic moment
operator made of pure gluon fields. We find that the leading
contribution of the matrix element is related to the nucleon's
tensor charge and is positive.
We thus conclude that a sea-quark contribution to the
proton's magnetic moment depends non-trivially on the sea-quark mass:
It must change sign at some critical value of the sea-quark mass. We argue that this critical value could be smaller than the physical mass of the strange quark.

\vskip 0.15cm {\bf 2.} {\it Effective gluonic operator from heavy
quark current.} Consider a proton of momentum $p$ and polarized in
the $z$-direction, $\pzr$. The $z$-component of its magnetic moment
is defined as
\beq \mu_p=\frac{\pzl \frac12 \int d^3\vec{r} \;
(\vec{r} \times \vec{j}_{\rm em})_z \pzr}{\pzn}\ ,
\eeq
where $\vec{j}_{\rm em}$ is the electromagnetic current.
In order to carry out the spatial integration, it is useful to
consider the off-diagonal matrix element and to take the forward
limit after integrating over $\vec{r}$ \beq \mu_p=-\frac{i}2
\left.\Big( \vec{\nabla}_{\vec{q}} \times \ppzl \;\vec{j}(0)\; \pzr
\Big)_z\right|_{\vec{q}=0} \ , \eeq where $q=p'-p$ is the momentum
transfer.

Now consider the contribution to the magnetic momentum from a heavy
quark with electromagnetic current $j^\mu_Q = {\overline
Q}\gamma^\mu Q$. (The quark also contributes to the magnetic moment
through wave functions. However, this is conventionally regarded as
a part of the light-quark contribution.) To lowest order in
$1/m_Q$, its contribution to the form factor is given by the diagram
shown in Fig.~1. When $m_Q\gg \Lambda_{\rm QCD}$, the quark loop can
be integrated out perturbatively. The heavy-quark electromagnetic
current is matched to a leading effective operator consisting of
pure gluon fields \cite{Kaplan:1988ku}
\beq
j_{\mu Q}= C(m_Q)\;
\partial^\alpha T_{\mu\alpha}(m_Q) + ... \ ,
\eeq
where ellipses denote
higher order contributions in $1/m_Q$, and $
C(m_Q)=g^3(m_Q)/((4\pi)^2 45 m_Q^4) $  to leading order in the
strong coupling $g(m_Q)$, and
\begin{equation}
T_{\mu\alpha}= 14 \Tr \;G_{\mu\sigma}
\{G^{\sigma\tau},G_{\tau\alpha}\} -5 \Tr \;G_{\sigma\tau}
\{G^{\sigma\tau},G_{\alpha\mu}\} \ ,
\end{equation}
where
$G_{\mu\nu}$ is the strong field-strength tensor.
Notice that the tensor $T_{\mu\alpha}$ is antisymmetric
and that
the effective gluon operator is renormalized at the scale $m_Q$.

\begin{figure}[h]
\hspace{-.4cm}\vspace{-.2cm}
\includegraphics[scale=0.4, clip=true, angle=0,
draft=false]{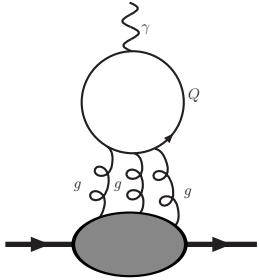} \caption{\label{fig1} Lowest order heavy
sea quark contribution to the proton's magnetic moment.}
\end{figure}

The matrix element of the heavy-quark current can thus be written as
\bea \ppzl j_{\mu Q} \pzr
=C(m_Q) \ppzl \partial^\alpha T_{\mu\alpha} \pzr \nn \\
= i C(m_Q) q^\alpha \pzl T_{\mu\alpha} \pzr \ .
\eea From Eq. (2), the heavy-quark contribution
to the proton's magnetic moment is then given by
\begin{equation} \label{muQ} \delta\mu_{p}^{Q}=C(m_Q) \; \pzl T_{yx} \pzr \ ,
\end{equation}
with
\begin{equation}
T_{yx}  = \sum_{abc} d^{abc} \Big[ \; 7(\vec{E}^a\cdot\vec{B}^b)
\; E^{cz} - 2 (\vec{E}^a\cdot\vec{E}^b-\vec{B}^a\cdot\vec{B}^b) \;
B^{cz} \Big] \ ,
\end{equation}
where $d^{abc}$ are the $SU(3)_{\rm c}$ coefficients, and
$\vec{E}^a$ and $\vec{B}^a$ are the chromo-electric and
chromo-magnetic fields, respectively.  We have followed the
convention in the literature which leaves out the electric charge,
$e_Q$, factor. As is obvious from Fig.~1, the effective operator is
exactly the same as that obtained for light-by-light scattering at
lowest order in quantum electrodynamics (QED) \cite{jackson}.

\vskip 0.15cm
{\bf 3.} {\it Top quark contribution to the magnetic moment
of the $\Lambda_b$ baryon.} To understand the nature of the gluonic
matrix element in the proton, let us consider a similar case in QED.
At order $\alpha_{\rm em}^3$, the electron's anomalous magnetic
moment receives a contribution from the muon through
light-by-light scattering. In the limit $m_\mu/m_e\gg 1$, this quantity has been analytically calculated in \cite{Laporta:1992pa}:
\begin{equation}
  \left. \frac{g_e-2}{2}\right|_{\rm from~\mu}
    =\left(\frac{\alpha_{\rm em}}{\pi}\right)^3
    \left(\frac{m_e}{m_\mu}\right)^2\left[\frac{3}{2}\zeta(3)
   -\frac{19}{16}\right] + ... \ .
\end{equation}
This is a striking result: The leading power dependence on the
muon mass is $1/m_\mu^2$, rather than $1/m_\mu^4$ as implied by a
naive heavy-$m_\mu$ expansion similar to (4-8). The discrepancy is resolved if one
realizes that the effective operator $T_{yx}^\gamma$ (obtained by
replacing the gluon fields by photon fields in Eq. (5)) is
renormalized at scale $m_\mu$. Power counting indicates that the
matrix element of $T_{yx}^\gamma$ in the electron state is
quadratically divergent, and the divergence is cut-off naturally by
the scale $m_\mu$. Since the matrix element of the effective operator
has dimension 3, it must behave as ($\langle e|e \rangle=(2\pi)^3\delta^3(0)$)
\begin{equation}
   \langle e|T_{yx}^\gamma|e\rangle \sim m_\mu^2 m_e \ ,
\end{equation}
where the $m_e$ dependence comes from the chirality flip of the
electron. This demonstrates that the matrix element of the effective
operator is dominated by quantum fluctuations at the scale of the heavy
fermion mass.

The QED result can be directly applied to an interesting case in
QCD. Consider a spin-1/2 $\Lambda_b$ baryon made of a heavy bottom
quark and two light up and down quarks coupled to zero spin and
isospin. To leading order in $1/m_b$, the spin of $\Lambda_b$ is
carried by the heavy $b$-quark. The magnetic moment of $\Lambda_b$
from the top quark sea can be calculated just like in the QED case.
Including a color factor of $C_d=\sum_{abc}d_{abc}^2/48 = 5/18$, the
leading order contribution in $\Lambda_{\rm QCD}/m_b$, $m_b/m_t$,
and $\alpha_s$ is
\begin{equation}
  \delta\mu^t_{\Lambda_b}
  =\mu_{\Lambda_b} \left(\frac{\alpha_{\rm s}}{\pi}\right)^3
   C_d \left(\frac{m_b}{m_t}\right)^2\left[\frac{3}{2}\zeta(3)
   -\frac{19}{16}\right] + ... \ ,
\end{equation}
where $\mu_{\Lambda_b}=e_t\hbar/2m_{\Lambda_b}c$. We thus conclude that the top quark contributes positively to the $\Lambda_b$ magnetic moment.

\vskip 0.15cm
{\bf 4.} {\it Gluonic matrix element in the proton}.
The above example is very instructive for the behavior of the
effective gluonic operator matrix element in the proton: Its
dependence on the renormalization scale $m_Q$ and $\Lambda_{\rm
QCD}$ must exhibit the form,
\begin{equation}
\label{pTyxp}
\pzl T_{yx} \pzr
   = a m_Q^2 \Lambda_{\rm QCD} + b\Lambda_{\rm QCD}^3 \ ,
\end{equation}
where the possible logarithmic dependence on $m_Q$ is neglected. The
$a$-term depends quadratically on $m_Q$ and is therefore dominant.
It receives contributions mainly from quantum fluctuations at scale
$m_Q$. Power counting of the diagrams involved in (\ref{pTyxp}) shows that this term comes from single quark contributions.  Diagrams involving two or three quarks from the proton cannot produce a term quadratic in $m_Q$. The $b$-term is entirely determined by physics at the
non-perturbative scale $\Lambda_{\rm QCD}$.

We first focus on the $a$-term since it dominates the matrix element (\ref{pTyxp}). Because it involves
contributions from both large momentum flow and scale $\Lambda_{\rm
QCD}$, we factorize them by matching the dimension-6 operator
$T^{\mu\nu}$ to a set of dimension-4 ones $\theta^{\mu\nu}_i$,
\begin{equation}
    T^{\mu\nu} = m_Q^2 \sum_i C_i \theta^{\mu\nu}_i + ... \ ,
\end{equation}
where $\theta^{\mu\nu}_i$ must be gauge-invariant second-order
antisymmetric tensors with odd charge parity, and ellipses denote
higher dimensional operators. Since we are interested in the forward
matrix element, total derivative operators are ignored. Such gauge invariant
dimension-4 operator cannot be constructed with pure gluonic fields. There are, however,
two such operators made of quark fields,
\begin{equation}
    \epsilon^{\mu\nu\alpha\beta}\overline{\psi}
       (iD_\alpha\gamma_\beta - iD_\beta\gamma_\alpha)\psi, ~~~
          m\overline{\psi}\sigma^{\mu\nu}\psi \ .
\end{equation}
Using the QCD equation of motion $(i\not\!\!D-m)\psi=0$, the first
operator can be reduced to the second one. Therefore, we find that
\begin{equation}
   T^{\mu\nu} = a m_Q^2\sum_f
   m_f\overline{\psi}_f\sigma^{\mu\nu}\psi_f
   + ... \ ,
\end{equation}
where $a = \frac12 (\alpha_s/\pi)^3 C_d[3/2\zeta(3)-19/16]$ comes from the
matching calculation for a single quark state, and the sum runs over
the light-quark flavors.


It is now straightforward to calculate the leading contribution to
the magnetic moment,
\begin{equation}
    \delta\mu_p^Q = C(m_Q) m_Q^2 a (m_u\delta u + m_d\delta d) + ... \ ,
\end{equation}
where $\delta u$ and $\delta d$ are the tensor charges of the proton
defined as $
    \langle PS|\overline{\psi}i\sigma^{\mu\nu}\gamma_5\psi|PS\rangle
     =2 \delta \psi (S^\mu P^\nu - S^\nu  P^\mu)$,
with $P$ and $S$ the proton's four-momentum and spin, respectively
\cite{Jaffe:1991kp}. The above equation is one of the main results
of this paper.  Using the PDG values $m_u = 2.8\pm 1.3$ MeV and $m_d
= 6\pm 2$ MeV \cite{Eidelman:2004wy} and the tensor charge from the
MIT bag model $\delta u=1.2$ and $\delta d= -0.3$
\cite{Jaffe:1991kp}, we find that $\delta\mu^Q_p$ is positive.
Lattice calculations at higher renormalization scales yield a
similar ratio $\delta u/\delta d$ \cite{Gockeler:2005cj} which does
not change the conclusion.

The strange quark also contributes to Eq.~(16) through
the proton wave function, but its contribution is suppressed compared to that of the light up and down quarks.  The scenario we study here is that the
strange quark is the fictitious heavy quark $Q$. The corresponding $m_Q\delta Q$
term in Eq.~(16) can be calculated by matching $\overline{Q}\sigma^{\mu\nu}Q$ to
the gluon operator $T^{\mu\nu}$ and then evaluating the leading
contribution from the matrix element $T^{\mu\nu}$ in the proton
state. The result is that although $m_Q\delta Q$ is a leading power,
it is suppressed by $\alpha_s^3(m_Q)$ relative to the up and down
quark contributions and is therefore significantly smaller.

\vskip 0.15cm
{\bf 5.} {\it An example of non-perturbative
contribution.} The subleading non-perturbative contribution is hard
to calculate without a specific model of the proton. Here we present
quark model calculations where the excitations of the
quarks and the non-linear gluonic interactions are neglected. In
this Abelian approximation, the gluon fields consist of eight copies
of the electromagnetic field obeying the linear Maxwell equations.
The chromo-electric fields are given by $
\vec{E}^a(\vec{r})= \frac{\lambda^a}2 \vec{E}(\vec{r}) = -
\frac{\lambda^a}2 \vec{\nabla} \Phi(\vec{r}), $ with \beq
\Phi(\vec{r})=\frac{g}{4\pi} \int d^3r_1
\frac{\rho(\vec{r}_1)}{|\vec{r}-\vec{r}_1|} \ , \eeq where the index
$a$ refers to its color,
$\rho(\vec{r})=\langle\alpha|\rho(\vec{r})|\beta\rangle$ denotes the
charge density matrix element between two states, $|\alpha\rangle$
and $|\beta\rangle$, and $\lambda^a$ are the Gell-Mann matrices.
Similarly, the chromo-magnetic fields are given by $
\vec{B}^a(\vec{r})= \frac{\lambda^a}2 \vec{B}(\vec{r}) =
\frac{\lambda^a}2 \vec{\nabla}\times\vec{A}(\vec{r})$ with
\beq
\vec{A}(\vec{r})=\frac{g}{4\pi} \int d^3r_1
\frac{\vec{j}(\vec{r}_1)}{|\vec{r}-\vec{r}_1|} \ ,
\eeq
where $\vec{j}(\vec{r})=\langle\alpha|\vec{j}(\vec{r})|\beta\rangle$ is
the quark current density matrix element.

We compute the contribution to the magnetic moment due to a single
quark polarized in the $+z$-direction, and is in the ground state of
a quark model potential.  The electric and magnetic fields can be calculated using the wave functions that
describe the quark fields in the ground state of the model. We
use the non-relativistic (NR) quark model and the MIT bag
model \cite{bag,bhaduri}. The dominant contribution to the proton
magnetic moment is \bea \label{muQsingle}
\delta\mu^Q_{q}&=&C(m_Q) \int dr \; r^2\;m(r) \ , \\
m(r)&=& 7\vec{E}_{\ua\ua} \cdot\vec{B}_{\ua\ua} \; E^{z}_{\ua\ua} -
2 \vec{E}_{\ua\ua}\cdot\vec{E}_{\ua\ua}\; B^{z}_{\ua\ua} \nn \\
&&\hspace{.0cm}+2 \Big(\vec{B}_{\ua\ua}\cdot\vec{B}_{\ua\ua} \;
B^{z}_{\ua\ua} +\frac13 \vec{B}_{\ua\da}\cdot\vec{B}_{\da\ua} \;
B^{z}_{\ua\ua}\nn\\
&&\hspace{.2cm}+\frac23 {\rm Re}
\vec{B}_{\ua\ua}\cdot\vec{B}_{\ua\da} \; B^{z}_{\da\ua} \Big) \ .\nn
\eea The numerical evaluation of these integrals results in
$\delta\mu_q^Q/\mu_NC(m_Q)\simeq2.2$~fm$^{-4}$ and $1.3$~fm$^{-4}$
in the NR quark model and in the MIT bag model, respectively. If
naively applying this result for the physical strange quark, with
$m_s=110$~MeV and $\alpha_s(m_s)\simeq1$, we get
$\delta\mu_p^s\simeq 0.1e_s\mu_N$, where $\mu_N=e\hbar/2Mc$ is the
nuclear magneton.

In both models, the $\delta\mu_p^Q$ is therefore positive.  However, this result is reached for different reasons.  In
the NR  model, the $\vec{B}$ field is generated by the color
dipole of the quark and hence is correlated with its spin. The
triple-$\vec{B}$ term can be neglected because the quark is
non-relativistic. When averaging over the direction of the $\vec{B}$
and $\vec{E}$ inside the proton, the term
$7(\vec{E}\cdot\vec{B})E^z$ dominates over $2\vec{E}^2B^z$. In the
MIT bag, on the other hand, the color current is generated from the
orbital motion of the quark. The direction of motion is correlated
with the spin. Here the triple-$\vec{B}$ term is non-negligible.
After canceling the negative $\vec{E}^2 B^z$ term, the result is
again positive.

It would be nice, of course, to also calculate contributions from
intermediate excited states, particularly the contribution with
excitation energy of order $m_Q$. Because of the intermediate state
sum, the single-quark matrix element dominates the $b$-term. Such a
calculation is beyond the scope of this paper. In the NR quark
model, it can be shown that all excited state contribution is
positive. For the MIT bag, on the other hand, we have only verified
this for the first few excited states.

\vskip 0.15cm {\bf 6.} {\it Mass-dependence of the sea quark
contribution to the proton's magnetic moment.} Thus it appears that
a heavy sea quark contributes positively to the
proton's magnetic moment. On the other hand, there are
indications, that a light sea quark contributes negatively.

One sensible picture for the light quark sea is the meson cloud model
\cite{Musolf:1993fu,Leinweber:2004tc}. If the strange quark is light,
the kaon is then a Goldstone boson. The proton has a
certain probability to dissociate into a kaon, which
contains an anti-strange quark, plus a $\Lambda$, which contains a strange quark.
The kaon carries a unit of orbital
angular momentum and its contribution to the magnetic moment is
negative. The $\Lambda$'s spin and magnetic moment are dominated by the
strange quark and the latter is positive. However, in the virtual
dissociation, the $\Lambda$ is polarized mainly against the proton
spin. Thus the strange quark contribution to the magnetic moment is
also negative. Therefore the total contribution from light strange and anti-strange
quarks is negative.

There is also an indirect lattice QCD calculation for the nucleon
sea contribution to the magnetic moment of the proton
\cite{Leinweber:2004tc},
\begin{equation}
   \delta \mu_p^{u-\rm loop} = \delta\mu^s_p - 0.3(1) \ ,
\end{equation}
where $\delta\mu_p^{u-\rm loop}$ is the up (or down) quark ``sea
contribution'' defined in the same sense as the strange quark sea.
Thus unless the real strange quark contribution is larger than
0.3$\mu_N$ (the recent world data average gives
$\delta\mu^s_p(Q^2\sim 0.1 {\rm GeV}^2)=(0.28\pm 0.20)e_s \mu_N$
\cite{happex}), the light-sea contribution shall be negative.

\begin{figure}[h]
\hspace{-.4cm}\vspace{-.0cm}
\includegraphics[scale=0.35, clip=true, angle=0]{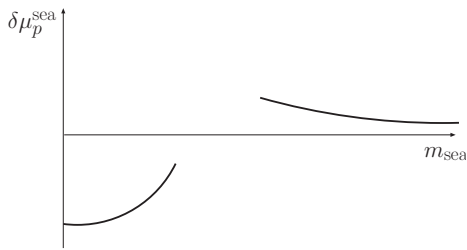}
\caption{\label{fig4} A possible picture for a sea-quark
contribution to the proton's magnetic moment as a function of the
sea-quark  mass.}
\end{figure}

The ineluctable consequence of these results obtained in the light
and heavy quark limits is that the sea-quark contribution to the
proton's magnetic moment vanishes at least once at some critical
value of the sea-quark mass.
How large is this critical mass? A plausible answer is that its
value is slightly above the masses of the up and down quarks, i.e.
around 10 to 20 MeV. Indeed, this is where the chiral behavior seems
to rapidly set in, as can be seen from chiral extrapolations of
lattice data \cite{Leinweber:2004tc}. If this is correct, the
strange quark contribution to the magnetic moment of the proton is
positive.

\vskip0.15cm {\bf 7.} {\it Conclusion.} In this paper, we advocate
the study of the contribution to the magnetic moment of the proton
from a heavy-quark sea. We show that the leading contribution can be
calculated in perturbative QCD combined with the quark tensor charge
and that the contribution is positive. If one believes that the
contribution is negative when the quark mass is small, the magnetic
moment has a nontrivial dependence on the sea-quark mass. Therefore
the sign of the strange quark contribution to the proton's magnetic
moment is very sensitive to the QCD dynamics.

The authors would like to thank D.~Beck, M.~Burkardt, A. Manohar,
M.~Ramsey-Musolf, and R. Young for discussions related to the
subject of this paper. The work of D.~T. is supported by NSF grant
NSF-PHY-0102409. He wishes to thank the Particle Physics Group at
the Rensselaer Polytechnic Institute, where part of this work was
completed.  X.~J. is supported by the U. S. Department of Energy via
grant DE-FG02-93ER-40762 and by a grant from the Chinese National
Natural Science Foundation (CNSF).

\end{document}